\documentclass[aps,prb,showpacs,reprint,amsmath,amssymb,groupaddresses,superscriptaddress]{revtex4-1}

\usepackage{graphicx}
\usepackage{dcolumn}
\usepackage{bm}

\begin{document}


\title{Optically induced nuclear spin polarization in a single GaAs/AlGaAs quantum well probed by a resistance detection method in the fractional quantum Hall regime}


\author{K. Akiba}
\affiliation{JST, ERATO Nuclear Spin Electronics Project, Sendai, 980-8578, Japan}

\author{T. Yuge}
\affiliation{Department of Physics, Osaka University, Machikaneyama-Cho, Toyonaka, 560-0043, Japan}

\author{S. Kanasugi}
\affiliation{Department of Physics, Tohoku University, Sendai, 980-8578, Japan}

\author{K. Nagase}
\affiliation{JST, ERATO Nuclear Spin Electronics Project, Sendai, 980-8578, Japan}

\author{Y. Hirayama}
\affiliation{JST, ERATO Nuclear Spin Electronics Project, Sendai, 980-8578, Japan}
\affiliation{Department of Physics, Tohoku University, Sendai, 980-8578, Japan}
\affiliation{WPI-AIMR, Tohoku University, Sendai, 980-0812, Japan}




\begin{abstract}
We study the optically pumped nuclear spin polarization 
in a single GaAs/AlGaAs quantum well in the quantum Hall system. 
We apply resistive detection via the contact hyperfine interaction, 
which provides high sensitivity and selectivity, 
to probe a small amount of polarized nuclear spins in a single well. 
The properties of the optical nuclear spin polarization are clearly observed. 
We theoretically 
discuss the nuclear spin dynamics accompanied with doped electrons 
to analyze the experimental data.
The optical nuclear polarization spectra exhibit
electron-spin-resolved lowest Landau level interband transitions. 
We find that
the phonon emission process, which normally assists the optical pumping process,  
influences the optical nuclear spin polarization. 
We also discuss 
that the electron-electron interaction can play an important role in the optical nuclear spin polarization.
\end{abstract}




\maketitle

\section{Introduction}
Optical nuclear polarization 
accomplishes 
the signal enhancement of nuclear magnetic resonance (NMR)\cite{OpticalOrientation}
and also promises to be requisite elements such as the initialization 
for the nuclear spin quantum information technology.\cite{SpinsQubits, Goto, Reimer}
In this method, i.e., optical pumping, 
the controllability of laser illumination enables us 
to manipulate the sign and magnitude of the nuclear spin polarization. 
To implement the quantum information processing, 
the coherent manipulation of nuclear spins is essential. 
The coherent nuclear spin operation has been utilized in the conventional NMR technique 
through the irradiation of the radio frequency (RF) magnetic field. 
The issue in the conventional method is the low sensitivity for small numbers of nuclear spins 
due to the small amount of nuclear spin polarization.

Recently, the control of multiple quantum coherences of nuclear spins in a nanometer-scale region 
has been demonstrated by using resistive detection in the quantum Hall system.\cite{YusaNature} 
This type of microscopic NMR technique, 
which is performed by the RF irradiation from a miniature antenna, 
is expected to be a good candidate 
to implement quantum information processing using nuclear spins as multiple qubits.
The resistive detection provides the high sensitivity to observe a small ensemble of nuclear spins. 
Since the optical pumping enables us to generate various nuclear spin polarizations, 
the resistively detected NMR in the quantum Hall system combined with the optical nuclear polarization 
opens up a new possibility of rich quantum information processing. 
It is important to determine 
the detailed properties of optical pumping 
by using resistive detection in the quantum Hall system 
in order to utilize this combined technique effectively for such processing.  

Optical nuclear polarization in a two-dimensional electron system (2DES) 
was first reported by Barrett {\it et al.},
where the GaAs/AlGaAs multiple quantum well and the conventional (coil-detection) NMR were used.\cite{Barrett}
Although this is a milestone for the optically pumped NMR (OPNMR) study for the quantum Hall system, 
a number of questions about the photophysics and spin physics of the optical pumping process 
were raised.\cite{TyckoReview}
Vitkalov {\it et al.} studied 
the dynamic nuclear polarization pumped by unpolarized light in multiple quantum wells 
near the Landau level filling factor $\nu=1$.\cite{Vitkalov} 
Moreover,  
the OPNMR   
revealed fascinating spin physics in the quantum Hall regimes, such as Skyrmions.
\cite{Tycko, Skyrmion, OPNMRaLocalProbe} 
Nevertheless, 
the properties of the optical nuclear polarization in the quantum Hall system 
have not been fully investigated.
Instead of the conventional detection for nuclear spins using a coil, 
Kukushkin {\it et al.} and Davis {\it et al.} utilized optical detection to observe 
the optical nuclear polarization in a single heterojunction.\cite{Kukushkin, Davis}
However, optical detection can probe only the local high nuclear polarization, 
which is different from the observation in conventional coil-detection.
Indeed, the obtained nuclear polarizations were much higher than that in Ref.~\onlinecite{Barrett}.  
Thus, the optical polarization of the nuclei interacting with 2DES 
in the quantum Hall regime 
has not yet been clearly elucidated. 

In this paper, 
we describe our investigation of the optical nuclear polarization in a single quantum well 
by using resistive detection. 
This detection allows us 
to selectively probe the nuclear spins 
interacting with the 2DES via the contact hyperfine coupling. 
Hence, the region carrying the conduction electrons is detected,  
and the signal is not affected by the nuclei outside the well. 
This is contrastive to the conventional NMR 
where all the nuclei in the sample are detected. 
Unlike in optical detection, the resistively detected signal 
is not limited to the optically induced local phenomena, 
even though the current-flowing region is not homogeneous in the well.  
The signal by the resistive detection is strong enough to be observed from even a single quantum well. 
In the experiments, we utilized the shift of the resistance peak at $\nu=2/3$ for the detection of  the optical nuclear spin polarization,\cite{OPRD} 
which enables us to estimate the value of the nuclear magnetic field. 
We obtain clear data on the properties of the optical nuclear polarization 
in the quantum Hall regime using this resistive detection. 
Furthermore, 
we make a theoretical formulation of  
the nuclear spin dynamics accompanied with the inherent electrons
to analyze the experimental data. 
We theoretically derive the steady state electron spin polarization 
in a modulation doped system under optical pumping, 
which is the source of the nuclear polarization dynamics. 
The obtained knowledge provides new insights into the optically pumped nuclear spin polarization 
and is of importance in future quantum information technology.

\section{Methods}\label{Methods}

The sample is a 30-$\mu$m wide and 100-$\mu$m long Hall bar, 
which was processed from a wafer containing a single 
18-nm GaAs/Al$_{0.33}$Ga$_{0.67}$As quantum well 
with one-side $\delta$-doped barrier layer.   
The electron density $n_s$ was controlled by the back gate voltage $V_{bg}$ using a Si-doped $n^+$-GaAs substrate as a gate electrode. 
The sample was cooled in a cryogen free $^3$He refrigerator down to a 320-mK base temperature. 
The electron mobility was 185~m$^2$/(Vs) for $n_s=1.2\times 10^{15}$~m$^{-2}$ after illumination. 
A mode-locked Ti:sapphire laser (pulse width: $\sim 2$~ps, pulse repetition:~76 MHz) 
was used for the optical pumping. 
The laser illumination was switched by using an acousto-optic modulator. 
A laser beam irradiated the whole Hall bar structure (beam diameter: 200~$\mu$m)
through an optical window on the bottom of the cryostat. 
The propagation direction of the laser beam was parallel to the external magnetic field $B=7.15$~T, 
which was applied perpendicular to the quantum well. 

The measurement procedure was as follows. 
First, the nuclear polarization was initialized by setting 
the electronic state to the Skyrmion region ($\nu=1.1$) by $V_{bg}$ 
for 80~s to depolarize the nuclear spins.\cite{Smet} 
Second, optical pumping was performed 
at $\nu\sim0.3$ 
for time duration $\tau_{pump}$ 
with the wavelength $\lambda$ and the average power density $P$. 
The laser illumination increased  
the temperature of the $^3$He pot,
which was thermally-connected to the sample, 
up to 380~mK; 
the illumination also increased the sample resistance.  
Third, $\nu$ was set to 1 for 70~s 
so that the resistance returned to the value before the illumination, 
where the relaxation of the nuclear polarization at $\nu=1$ was the smallest 
within the available $V_{bg}$.  
Finally, we measured the longitudinal resistance $R_{xx}$ 
by sweeping up $V_{bg}$ around $\nu=2/3$, 
using a lock-in technique and a 30-nA sinusoidal current ($79$~Hz).\cite{noteDNP} 

The resistive detection is achieved by 
the change in the electron Zeeman energy induced by 
the nuclear polarization via the contact hyperfine interaction.
We employed the spin phase transition (SPT) between the polarized 
and unpolarized  
phases
at $\nu=2/3$ as the phenomenon sensitive to this change. \cite{OPRD} 
This SPT occurs at a level crossing point 
where 
the composite fermion cyclotron energy $E_c=c_1 \sqrt{B} (\nu-1/2)$ coincides with the Zeeman energy $E_Z=c_2 (B+B_N)$. 
Here, $c_1$ and $c_2$ are constants, and $B_N$ is the nuclear magnetic field. 
The preferable
phase is an unpolarized and polarized state for $E_c>E_Z$
and $E_c<E_Z$, respectively.
Since $\nu$ is a linear function of $V_{bg}$ in a fixed magnetic field, 
the SPT is observed as the $R_{xx}$ peak by sweeping $V_{bg}$ and 
its position $V_{peak}$ ($V_{bg}$ at the peak) depends on $B_N$. 
From these, we derive $B_N=c\, \Delta V_{peak}/\sqrt{B}$, 
where $\Delta V_{peak}$ is defined as the peak shift from $V_{peak}$ with $B_N=0$
and $c$ is a proportional constant. 
$c$ was determined from the peak coincidence measurement \cite{OPRD} 
to be 80.6~V$^{-1}$T$^{3/2}$ in our sample. 
Thus, we obtained the nuclear magnetic field $B_N$ by optical pumping 
from a shift of the SPT peak.\cite{Note1}

\section{Theory}\label{Theory}

The cross relaxation between electron spins of the 2DES and nuclear spins 
is dominated by the fluctuations of 
the Fermi contact (scalar) hyperfine interaction.\cite{Vitkalov, Dobers} 
Assuming that the optical pumping is homogeneous in the plane parallel to the well ($xy$-plane),  
we can describe the temporal evolution of the nuclear spin polarization 
$\langle I_z (z,t) \rangle$ along the magnetic field direction ($z$-direction) as\cite{Tycko, Abragam} 
\begin{align}
\frac{\partial \langle I_z(z,t)\rangle}{\partial t}= &D\frac{\partial^2 \langle I_z(z,t)\rangle}{\partial z^2}-\frac{1}{T_{IS}(z)} \biggl\{\langle I_z(z,t) \rangle - \langle I_z \rangle _{eq}\nonumber \\
&-\frac{I(I+1)}{S(S+1)}\left(\langle S_z \rangle-\langle S_z \rangle _{eq}\right)\biggr\}. 
\label{master}
\end{align}
$D$ is the nuclear spin diffusion constant; 
$T_{IS}(z)$ is the spatially dependent electron-nuclear cross relaxation time;
$\langle I_z \rangle_{eq}$ is the thermal equilibrium nuclear spin polarization; 
$S\ (=1/2)$ and $I$ are the electron and nuclear spin quantum numbers, respectively; 
$\langle S_z \rangle$ is the electron spin polarization;
and $\langle S_z \rangle_{eq}$ is the thermal equilibrium electron spin polarization.  
The values of $\langle I_z \rangle_{eq}$  for all nuclear species are less than 1\% 
under the experimental condition,\cite{ReviewReimer} and thus, we can neglect $\langle I_z \rangle_{eq}$. 
Here, we discuss the influence of the spin diffusion on the nuclear polarization in the well. 
The nuclear spin diffusion is driven by the nuclear magnetic dipole-dipole interaction 
between the same nuclear species.  
The value of $D$ in the AlGaAs barrier should be different from that in the GaAs well; 
Indeed, $D \sim 10^{-13}$~cm$^2$/s in GaAs  
and $D \sim 10^{-14}$~cm$^2$/s in Al$_{0.35}$Ga$_{0.65}$As have been reported.\cite{diffusionGaAs, diffusionAlGaAs} 
It has also been reported 
that nuclear spin diffusion is strongly suppressed at the GaAs/Al$_{0.33}$Ga$_{0.67}$As interface
and the effective $D$ in GaAs is 10$^{-15}$~cm$^2$/s. \cite{diffusioninterface}
Furthermore, in Ref.~\onlinecite{Tycko}, Tycko {\it et al.} mentioned that
the effective diffusion constant $D$ for the nuclear spin diffusion 
between the GaAs wells and the Al$_{0.1}$Ga$_{0.9}$As barriers 
is less than 10$^{-15}$~cm$^2$/s for the $\nu$-range from 0.6 to 1.8 in the quantum Hall system. 
Although the Al composition in our sample is 0.33 
and the value of $D$ is expected to be smaller than their values, 
we use $D=10^{-15}$~cm$^2$/s 
to roughly estimate the influence of the nuclear spin diffusion on the nuclear polarization in the well.    
Since the optical pumping preferentially creates nuclear spin polarization 
around the center of the well due to the spatial distribution of the electron density along the $z$-direction, 
the diffusion of the nuclear polarization to the outside of the well is not crucial
for $\tau_{pump}<(w/2)^2/D=810$~s, where $w$ is the width of the well.
In such a short pumping time, we can neglect the effect of the nuclear spin diffusion 
and obtain the solution of Eq. (\ref{master}): 
\begin{align}
{\langle I_z (z,t)\rangle}=&\frac{I(I+1)}{S(S+1)}\left({\langle S_z \rangle}-\langle S_z \rangle _{eq}\right) \nonumber \\
&\times\left\{1-\exp\left(-\frac{t}{T_{IS}(z)}\right)\right\}. 
\label{solutioneq}
\end{align}
The nuclear magnetic field experienced by the 2D electrons is of the form\cite{diffusionGaAs, Paget}
\begin{align}
B_N(t)=b_N\int \rho(z)\frac{\langle I_z (z,t)\rangle}{I} dz, \label{BNInt}
\end{align}
where $b_N$ is the full polarization nuclear magnetic field, 
and $\rho (z)$ is the conduction electron density envelope function. 
In our situation, we sum the contributions to the nuclear magnetic field from three nuclear species.  
The values of $b_N$ for $^{69}$Ga, $^{71}$Ga, and $^{75}$As are $-1.37$~T, $-1.17$~T, and $-2.79$~T, respectively.\cite{SpinPhysics} 
The negative sign of these values is due to the reduced electron $g$-factor in GaAs ($g^\ast =-0.44 < 0$). 
We note that $\rho (z)$ has large values around the center of the well,  
and that the same tendency is expected for $1/T_{IS}(z)$ 
because the electron-nuclear cross relaxation is induced by the contact hyperfine interaction.
Therefore, the integration in Eq. (\ref{BNInt}) is mainly dominated by the contribution from the center of the well.    
Since the hyperfine coupling constants for three nuclear species are in the same range,  
we assume the effective nuclear magnetic field takes on the following form:   
\begin{align}
B_N (t) = -A \left({\langle S_z \rangle}-\langle S_z \rangle _{eq}\right)\left\{1-\exp\left(-\frac{t}{T'_{IS}}\right)\right\}, \label{effectiveBN}
\end{align}
where $A\ (>0)$ is the constant, 
and $T'_{IS}$ is the effective electron-nuclear cross relaxation time.  
We use this equation (\ref{effectiveBN}) in the analysis. 

We should incorporate the effect of the doped electrons into $\langle S_z \rangle$ 
because they exist in the well under a thermal equilibrium condition without optical pumping. 
In order to obtain the steady state electron spin polarization in such a situation, 
we consider the following rate equations for numbers of up ($n_\uparrow$) and down ($n_\downarrow$) spin electrons 
with optical pumping:
\begin{align}
\frac{d n_\uparrow}{dt}&=G p_{\uparrow}-w_{\uparrow\downarrow}n_\uparrow +w_{\downarrow\uparrow}n_\downarrow -\frac{1}{\tau_e}\left(n_\uparrow-n_\uparrow^{eq}\right), \label{rateup} \\
\frac{d n_\downarrow}{dt}&=G p_{\downarrow}+w_{\uparrow\downarrow}n_\uparrow -w_{\downarrow\uparrow}n_\downarrow -\frac{1}{\tau_e}\left(n_\downarrow-n_\downarrow^{eq}\right), \label{ratedown}
\end{align}
where $G$ is the number of photo-generated electrons per unit time, 
$p_\uparrow$ $(p_\downarrow)$ is the up (down) spin excitation probability, 
$w_{\uparrow\downarrow}$ $(w_{\downarrow\uparrow})$ is 
the transition probability from up (down) to down (up) spins, 
$\tau_e$ is the electron lifetime (electron recombination time), 
and $n_\uparrow^{eq}$ $(n_\downarrow^{eq})$ is 
the number of thermal equilibrium up (down) spin electrons. 
Assuming that $w_{\uparrow\downarrow}$ $(w_{\downarrow\uparrow})$ 
is the same as that without optical pumping, 
we obtain the relationship 
$w_{\uparrow\downarrow}n_{\uparrow}^{eq}=w_{\downarrow\uparrow}n_{\downarrow}^{eq}$. 
We solve the steady state equations by using this relationship and obtain 
\begin{align}
{\langle S_z \rangle}
=&S\frac{n_{\uparrow}-n_\downarrow}{n_{\uparrow}+n_\downarrow}\\
=&\frac{n_{l}}{n_{eq}+n_{l}}\left(\frac{\langle S_z \rangle_{0}}{1+{\tau_{e}}/{T_{1e}}}+\frac{\langle S_z \rangle_{eq}}{1+{T_{1e}}/{\tau_e}}\right) \nonumber\\
&+\frac{n_{eq}}{n_{eq}+n_{l}}\langle S_z \rangle_{eq}, \label{steadyelectron}
\end{align}
where $n_{eq}=n_\uparrow^{eq}+n_\downarrow^{eq}$ is the number of thermal equilibrium electrons, 
$n_l=G\tau_e$ is the number of steady state photo-generated electrons, 
$T_{1e}=1/(w_{\uparrow\downarrow}+w_{\downarrow\uparrow})$ is the electron spin-lattice time (electron spin relaxation time), 
$\langle S_z \rangle_{0}=S(p_\uparrow -p_\downarrow)$ is the initially photo-generated electron spin polarization, 
and $\langle S_z \rangle_{eq}$ is the thermal equilibrium electron spin polarization. 
In the limit of $n_{eq}/n_l\to 0$, 
Eq. (\ref{steadyelectron}) becomes the steady state electron spin polarization 
in the previous study.\cite{ReviewReimer} 

The sign of $B_N$ is determined by the factor ${\langle S_z \rangle}-{\langle S_z \rangle}_{eq}$ (see Eq. (\ref{effectiveBN})). By substituting Eq. (\ref{steadyelectron}) in this factor, we obtain
\begin{align}
{\langle S_z \rangle}-{\langle S_z \rangle}_{eq}=\frac{n_{l}}{n_{eq}+n_{l}} \frac{\langle S_z \rangle_{0}-\langle S_z \rangle_{eq}}{1+\tau_e/T_{1e}}. \label{DSz}
\end{align}
Therefore, $\langle S_z \rangle_{0}-\langle S_z \rangle_{eq}\equiv \Delta S_z>0$ yields $B_N<0$ 
and
$\Delta S_z<0$ yields $B_N>0$. 
The photo-generated electron spin polarization $\langle S_z \rangle_{0}$ directly corresponds to the light helicity (polarization) based on the optical selection rule, 
and 
the thermal equilibrium electron spin polarization $\langle S_z \rangle_{eq}$ is $\nu$-dependent.

\section{results and discussion}
\begin{figure}
\includegraphics[width=1\columnwidth]{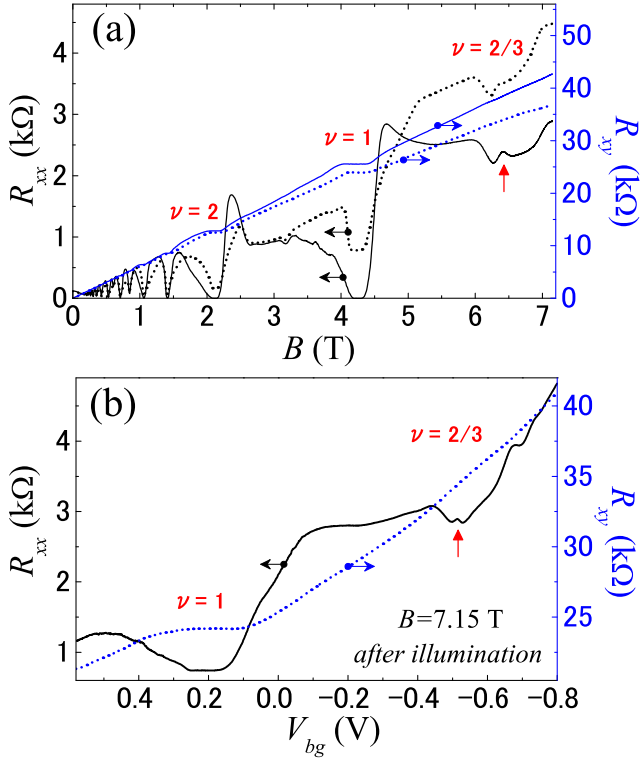}%
\caption{(Color online)
(a) $R_{xx}$ and $R_{xy}$ as a function of magnetic field with the carrier density of $\sim 1\times10^{15}$~m$^{-2}$. 
The solid (dotted) curves are $R_{xx}$ and $R_{xy}$ before (after) illumination. 
(b) $R_{xx}$ (solid curve) and $R_{xy}$ (dotted curve) at $B=7.15$~T obtained by sweeping $V_{bg}$.
\label{Fig:Resistance}}
\end{figure}

At first, we introduce the electronic transport properties of the sample we used. 
Figure~\ref{Fig:Resistance}~(a) shows 
$R_{xx}$ (left axis) and Hall resistance $R_{xy}$ (right axis) as a function of $B$. 
The solid and dotted lines indicate 
the resistances before and after illumination, respectively.  
In both cases, $n_s$ was tuned to $\sim 1\times10^{15}$~m$^{-2}$ by $V_{bg}$. 
In the solid lines (before illumination), 
we clearly observe the quantization of $R_{xy}$ ($\nu=1, 2$) accompanied by the $R_{xx}$ vanishment. 
The $R_{xx}$ dip appearing at $B$ from 6~T to 7~T originates from $\nu=2/3$ quantum Hall effect
and the peak at $B=6.4$~T (indicated by the upward arrow) 
is due to the SPT as mentioned in Sec.~\ref{Methods}. 
This SPT and the relatively high temperature (320~mK) causes 
no observation of the $R_{xx}$ vanishment and $R_{xy}$ quantization at $\nu=2/3$, 
since the peak becomes large and broad with increasing temperature.\cite{Kraus}
Once the sample was illuminated, 
the transport properties somehow changed without any illumination as shown in dotted lines. 
The quantization values of $R_{xy}$ decreases and $R_{xx}$ no longer vanishes.
These changes are due to the parallel conduction, which was created in the modulation doped layer after illumination.\cite{lowDsemicon,paracon} 
Despite the existence of the parallel conduction, 
we can also observe the SPT from $R_{xx}$.\cite{check} 
Figure~\ref{Fig:Resistance}~(b) shows $R_{xx}$ (left axis) and $R_{xy}$ (right axis) 
as a function of $V_{bg}$ at $B=7.15$~T after illumination. 
The $\nu=1$ integer quantum Hall effect and the SPT at $\nu=2/3$ can be observed
by changing the electron density at a constant magnetic field.
$V_{peak}$ is indicated by the arrow and the value of it is $-0.513$~V at $B=7.15$~T with $B_N=0$. 

\begin{figure}
\includegraphics[width=1\columnwidth]{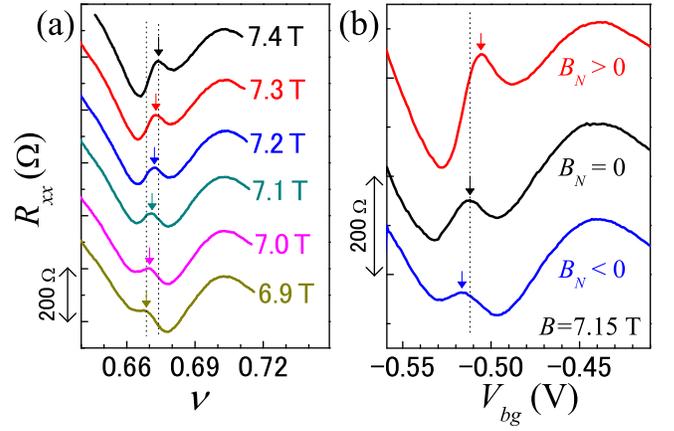}%
\caption{(Color online)
(a) The SPT peak in $V_{bg}$-sweep at several constant $B$-fields with $B_N=0$ after illumination, 
where $V_{bg}$ is converted to $\nu$ in the horizontal axis. 
(b) Examples of the SPT peak shift due to the existence of $B_N$ at $B=7.15$~T. 
The arrows indicate the peak positions. The curves are offset vertically for clarity.
\label{Fig:SPT}}
\end{figure}

Figure~\ref{Fig:SPT}~(a) shows the SPT peak in $V_{bg}$-sweep 
at several constant $B$-fields with $B_N=0$ after illumination, 
where $V_{bg}$ is converted to $\nu$ in the horizontal axis and the curves are offset vertically for clarity. 
The arrows indicate the peak positions. 
The peak position shifts to the higher $\nu$-side 
as $B$ is set to the larger value, 
where the variation of $R_{xx}$ as a function of $\nu$ is almost the same except the peak. 
This behavior makes us identify the peak as arising from the SPT.
As explained in Sec.~\ref{Methods}, 
the peak position (SPT point) is determined by $E_c=E_Z$. 
While $E_Z$ is proportional to $B$, $E_c$ is proportional to $\sqrt{B}$. 
Thus, when $B$ is increased, the SPT occurs at the higher $\nu$-side 
to satisfy $E_c=E_Z$, because $E_c$ linearly depends on $\nu$. 
Figure~\ref{Fig:SPT}~(b) shows 
examples of the SPT peak shift caused by the optical nuclear spin polarization at $B=7.15$~T. 
The arrows indicate $V_{peak}$ and the curves are offset vertically for clarity.
The upper and lower curves were obtained with $B_N=0.21$~T and $B_N=-0.15$~T, respectively.\cite{jouken} 
The middle curve was obtained without optical pumping (i.e., with $B_N=0$~T). 
In Fig.~\ref{Fig:SPT}~(b), $E_Z$ is modified by $B_N$ instead of $B$, 
and we can observe that the behavior of the peak shift is similar to that in Fig.~\ref{Fig:SPT}~(a).
It is noted that since $B_N$ only modifies $E_Z$ and does not affect $E_c$,  
the peak shift is attributed to the optical nuclear spin polarization 
and we can measure $B_N$ from the peak shift.

\begin{figure}
\includegraphics[width=1\columnwidth]{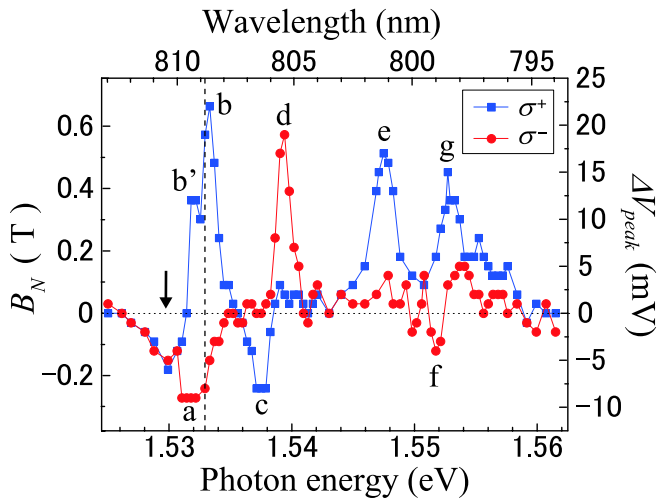}%
\caption{(Color online)
Photon energy dependence of optical nuclear polarization 
with $P=1.6$~W/cm$^2$ and $\tau_{pump}=150$~s. 
The squares and circles represent the $\sigma^+$ and $\sigma^-$ excitation, respectively. 
\label{Wavelength}}
\end{figure}

As explained in Sec.~\ref{Methods}, we measured the SPT peak shift with $\lambda$, $P$, and $\tau_{pump}$, 
and converted it into $B_N$. 
We obtain the properties of the optical nuclear spin polarization 
by varying the parameters and repeating the measurement. 

Figure \ref{Wavelength} shows the dependence of $B_N$ 
on the photon energy (laser wavelength $\lambda$) 
for both the $\sigma^+$ and $\sigma^-$ excitations, where $P=1.6$~W/cm$^2$ and $\tau_{pump}=150$~s.  
The excitation laser linewidth is $\sim$1~meV. 
We observe the clear peaks indicated by (a)--(g). 
To begin with, we take into consideration 
Eqs.~(\ref{effectiveBN}) and (\ref{DSz}) in Sec.~\ref{Theory}. 
With increasing $n_l$, the magnitude of $|B_N|$ increases and saturates at a fixed $\Delta S_z$, 
and thus, the observed spectra should depend on the photon absorption rate. 
The lowest energy peak (a) for $\sigma^{-}$ excitation (the peak (b) for $\sigma^{+}$ excitation) 
is located at 1.5318~eV (1.5333~eV).  
In a separate experiment, which was performed by using the same sample under the same conditions, 
the photoluminescence spectra show negatively charged exciton peaks 
that are relevant to the heavy-hole (HH) states,
and the lowest luminescence peak energies are 1.5310~eV and 1.5315~eV 
for the $\sigma^{-}$ and $\sigma^{+}$ polarizations, respectively. 
In the previous studies,\cite{Absorption, BJReview, Binding} 
the charged exciton and neutral exciton peaks appeared in the absorption spectra, 
and the charged exciton binding energy is less than 2~meV, which is comparable to the laser linewidth. 
Therefore, at least, 
the lowest energy peak (a) for the $\sigma^{-}$ excitation (peak (b) for $\sigma^{+}$ excitation)  
corresponds to an absorption process from the HH band into the lowest electron Landau level (LL$_0$). 
For (b), we observe the reproducible peak (b') located at 1.5320~eV. 
There is a possibility that (b') is the charged exciton peak and (b) is the neutral exciton peak. 
The second lowest energy peak (c) for the $\sigma^{+}$ excitation (peak (d) for $\sigma^{-}$) 
corresponds to the transition from the light-hole (LH) band to the LL$_0$. 
Peak (c) is $\sim 4$~meV above peak (b) and peak (d) is $\sim 8$~meV above peak (a). 
There is no discrepancy between the energy scale of these peak separations 
and that from the previous investigation.\cite{Absorption} 
The difference in the HH (LH) transition between $\sigma^{+}$ and $\sigma^{-}$  
is the excitation of the electron spin. 
By taking the optical selection rule into account, we assign the peaks to the transitions as follows.  
(a): HH with angular momentum $J_z=3/2$ $\Longrightarrow $ LL$_0$ with $S_z=1/2$; 
(b) and (b'): HH with $J_z=-3/2$ $\Longrightarrow $ LL$_0$ with $S_z=-1/2$; 
(c): LH with $J_z=-1/2$ $\Longrightarrow $ LL$_0$ with $S_z=1/2$; 
(d): LH with $J_z=1/2$ $\Longrightarrow $ LL$_0$ with $S_z=-1/2$. 
The sign of $B_N$ is consistent with these assignments 
because the excitation of the lower Zeeman level ($S_z=1/2$) leads to $\Delta S_z >0\ (B_N<0)$ 
and the excitation of the upper Zeeman level ($S_z=-1/2$) leads to $\Delta S_z <0\ (B_N>0)$, 
where we assume $|\langle S_z \rangle _{eq}|\neq 1/2$ based on our experimental conditions.\cite{SpinPol} 

The energy separation of 15~meV between (b) and (e) 
corresponds to the sum of the electron and HH Landau level splittings, 
where the Landau energy separation is 12~meV for the electrons and 2.3~meV for the HH 
when using effective mass $m_e^\ast=0.067m_e$ for the electrons and $m_{hh}^\ast=0.35m_e$ for the HH
($m_e$: mass of free electrons). 
Thus, we interpret that the higher energy peaks (e)--(g) correspond to  
the absorption processes reflected in the second Landau levels (LL$_1$).
However, it is difficult to assign a detailed transition between the electron and hole higher Landau levels. 
The hole complexity and the Fermi edge singularity\cite{FES} 
give rise to the difficulty of determining an absorption line. 
Since the relaxation in the higher levels includes the processes without nuclear-spin flip 
such as the cyclotron emission \cite{CE} and the Auger process,\cite{Auger} 
these processes diminish the magnitude of the polarization 
and the subsequent polarization can change the sign due to the electron-spin flip in these processes.
The negative nuclear magnetic field 
observed below 1.531~eV indicated by the arrow in Fig.~\ref{Wavelength} for both $\sigma^+$ and $\sigma^-$ excitation
is not fully understood. 
The localized electrons trapped by impurities may account for this helicity independent behavior. 

In the following, 
we focus on the lowest energy transition (HH $\Longrightarrow$ LL$_0$) at $\lambda=808.8$~nm (1.5329~eV ) indicated by the broken line in Fig.~\ref{Wavelength}
and investigate the properties of the optical nuclear polarization in the quantum Hall regime. 

\begin{figure}
\includegraphics[width=1\columnwidth]{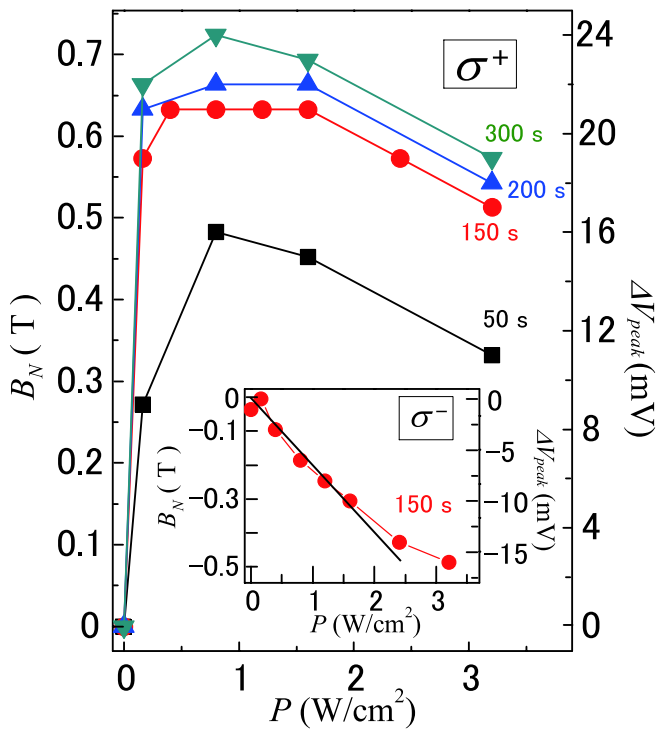}%
\caption{(Color online)
Power dependence of optical nuclear polarization 
for $\tau_{pump}=50~$s (squares), $\tau_{pump}=150~$s (circles), $\tau_{pump}=200~$s (triangles), and $\tau_{pump}=300~$s (inverted triangles) with $\sigma^+$ light. 
The inset shows the power dependence of the optical nuclear polarization for $\tau_{pump}=150~$s with $\sigma^-$ light.
\label{Intensity}}
\end{figure}

Figure \ref{Intensity} shows the dependence of $B_N$ on the laser power $P$ 
for various values of $\tau_{pump}$ with $\sigma^+$ light. 
With increasing $P$, the nuclear polarization sharply increases within the low $P$ range  
and gradually decreases above $P=0.8~$W/cm$^2$.
This behavior is almost independent of the illumination time. 
In Eq.~(\ref{DSz}), the prefactor $n_l/(n_{eq}+n_l)$ increases and saturates as $P$ increases,  
since $n_l$ is proportional to $P$. 
However, this does not completely account for the experimentally observed behavior of $B_N$
since $n_l$ is usually much less than $n_{eq}$. 
Indeed, the rough estimation gives $n_l=1\times 10^{12}~$m$^{-2} \ll n_{eq}=5\times 10^{14}~$m$^{-2}$, 
where $P=3~$W/cm$^2$, the absorption coefficient $\alpha=0.01$,\cite{FES} 
$\tau_e=1~$ns,\cite{Lifetime} 
$\lambda=808.8~$nm, $\nu=0.3$, and $B=7.15~$T.  
In such a range of $n_l$ (or $P$), $B_N$ is proportional to $P$, 
which explains only the initial increase in $B_N$ shown in Fig. \ref{Intensity}. 
We, therefore, take into consideration 
the other part $(\langle S_z\rangle_0-\langle S_z\rangle_{eq})/(1+\tau_e/T_{1e})$ in Eq. (\ref{DSz}) 
to explain the observed gradual decrease.

When the nuclear spins are polarized by the optical pumping, 
the energy difference between the electron and nuclear spins is usually compensated 
for by phonons.\cite{OpticalOrientation}
$\sigma^+$ light creates the electrons in the upper Zeeman level. 
In a simple picture, the electron relaxation from the upper to lower Zeeman levels induces the nuclear polarization. 
This process is accompanied by the phonon emission, and thus, causes the electron temperature to increase with increasing $P$. 
First, we consider the factor 
$\langle S_z\rangle_0-\langle S_z\rangle_{eq}\ (=\Delta S_z)$.
The value of $\langle S_z\rangle_{eq}$ is reported to be around 0.45 
at $\nu\sim 0.3$,\cite{SpinPol, Kuzma} where optical pumping was performed in our experiments. 
With increasing electron temperature, $\langle S_z\rangle_{eq}$ decreases to zero, 
and $\Delta S_z\ (=\langle S_z\rangle_0-\langle S_z\rangle_{eq})$ varies from $-0.95$ to $-0.5$, 
where $\langle S_z\rangle_{0}=-1/2$ for the $\sigma^{+}$ excitation. 
Second, we consider the factor $1/(1+\tau_e/T_{1e})$. 
In Ref.~\onlinecite{tT}, the temperature dependence of the electron spin relaxation in high magnetic field was investigated
and the ratio $\tau_e/T_{1e}$ increased with increasing temperature.
From this fact, 
we expect that $1/(1+\tau_e/T_{1e})$ decreases with the increase in temperature. 
Thus, 
when the heating power due to the phonon emission 
is larger than the cooling power of the cryostat, 
the change in $(\langle S_z\rangle_0-\langle S_z\rangle_{eq})/(1+\tau_e/T_{1e})$ 
is crucial for the nuclear spin polarization.\cite{Temp}  
We conclude that 
the suppression and decrease in $B_N$ in Fig. \ref{Intensity} 
is caused by the high rate of phonon emission.

To further confirm the influence of the phonon emission, 
we measure the optical nuclear polarization with $\sigma^-$ excitation, 
because $\sigma^-$ light creates electrons in the lowest energy level (the lower Zeeman level)
and the phonon emission process is not expected to polarize the nuclear spins. 
The inset in Fig. \ref{Intensity} shows  
the dependence of $B_N$ on $P$ for $\tau_{pump}=150~$s with $\sigma^-$ light. 
The magnitude of $|B_N|$ almost linearly increases. 
Thus, the variation in $B_N$ is dominated by $n_l$. 
The deviation from the linear function is understood due to the laser heating 
that increased the temperature of the $^3$He pot (the entire system). 
Therefore, the temperature increase we considered for the $\sigma^+$ illumination 
is qualitatively different from the laser heating effect in $\sigma^-$ illumination.  

We here discuss how nuclear spins are polarized with $\sigma^-$ illumination.
In a simple picture, 
in order to polarize nuclear spins, the electron-spin flip should occur 
from the lower to upper Zeeman levels, which requires excitation energy. 
Since the Zeeman energy gap corresponds to around 2~K 
and the temperature is less than 380~mK in our experiments,  
the thermal excitation probability of electrons should be quite low. 
However, surprisingly, we observed a negative $B_N$, 
which implies that there is another energy source in this process. 
The excitation energy can be provided by the electron-electron interaction. 
In our experiments, 
the Coulomb energy is not negligible compared with the Zeeman energy.\cite{Coulomb}
In such a case, 
even when only the lowest level (LL$_0$ with up spin) is occupied by electrons ($\nu<1$), 
the electron spin polarization does not always achieve a full polarization
due to the electron-electron interaction.\cite{SpinPol, Kuzma}  
This means that the electron-electron interaction can flip the electron spin.
In our analysis, we already included the effect of the electron-electron interaction 
through the value of $\langle S_z\rangle_{eq}\sim 0.45$. 
When we choose $\langle S_z\rangle_{eq}=1/2$ in a single particle picture, 
we obtain $B_N=0$ in the $\sigma^-$ excitation from Eqs.~(\ref{effectiveBN}) and (\ref{DSz}), 
that is, the nuclear spins cannot be polarized. 
Therefore, the electron-electron interaction can play an important role in optical nuclear spin polarization.

\begin{figure}
\includegraphics[width=1\columnwidth]{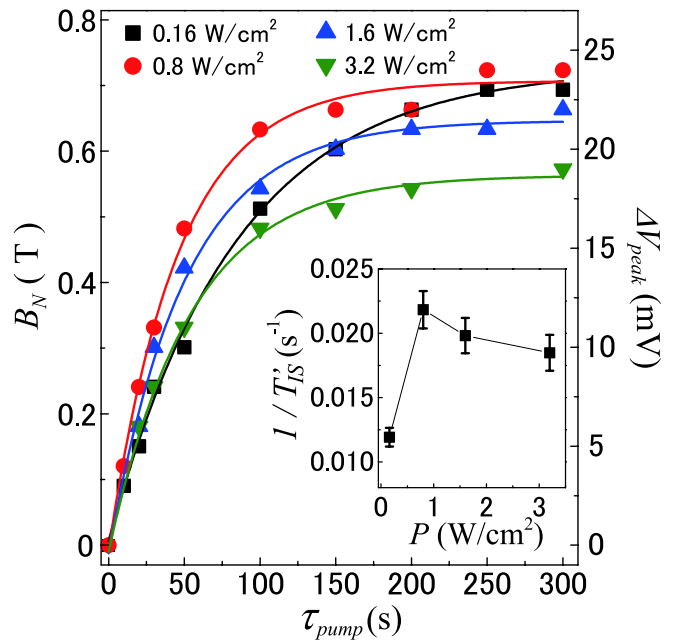}%
\caption{(Color online)
Illumination time dependence of optical nuclear polarization 
for $P=0.16~$W/cm$^2$ (squares), $0.8~$W/cm$^2$ (circles), 
$1.6~$W/cm$^2$ (triangles), and $3.2~$W/cm$^2$ (inverted triangles) 
with $\sigma^+$ light.
The solid lines are the fitting curves, as explained in the text.  
The inset shows the laser power dependence of the effective electron-nuclear cross relaxation rate. 
\label{Illuminationtime}}
\end{figure}

Figure \ref{Illuminationtime} shows the dependence of $B_N$ on $\tau_{pump}$   
for various $P$ values with $\sigma^+$ light. 
We fitted the data to Eq.~(\ref{effectiveBN}) 
and we obtain an effective electron-nuclear cross relaxation time $T'_{IS}$.  
The solid lines are the fitting curves.  
We obtain $T'_{IS}$ for $P=$0.16 W/cm$^2$, 0.8 W/cm$^2$, 1.6 W/cm$^2$, and 3.2 W/cm$^2$ of 
84 $\pm 5~$s, 46 $\pm 3~$s, 50 $\pm 3~$s, and 54 $\pm 4~$s, respectively. 
We find that 
the effective electron-nuclear cross relaxation rate $1/T'_{IS}$
depends on $P$, which is indicated in the inset in Fig.~\ref{Illuminationtime}. 
When $P$ increases above 0.8~W/cm$^2$, the electron temperature is expected to increase, 
as described in the explanation for Fig.~\ref{Intensity}. 
In this situation, if the lattice temperature also increases, 
the phonon emission rate is large. 
Since the phonon emission process is required to polarize nuclear spins in the $\sigma^+$ excitation, 
the large phonon emission rate results in a large electron-nuclear cross relaxation rate. 
This can account for the behavior in the inset. 

Another possibility for the change in $T'_{IS}$ is the nuclear spin diffusion effect. 
Although the NMR frequencies in the well 
are lower than that in the barriers due to the Knight shift, 
the laser irradiation forces the frequencies in the well to shift to the higher side 
because $\langle S_z\rangle_{eq}$ decreases with increasing $P$ as mentioned above.
The NMR spectrum in the well overlaps that in the barriers during illumination. 
Therefore, 
the energy matching enhances the spin flip-flop process in 
the nuclear magnetic dipole-dipole coupling between the well and the barriers,  
and the nuclear spin diffusion is thus accelerated. 
This acceleration may not allow us to neglect the nuclear spin diffusion effect, 
and the value of $T'_{IS}$ can be underestimated by
neglecting the diffusion term in Eq. (\ref{master}). 

\section{Conclusion}
We investigated the optically pumped nuclear spin polarization in a single GaAs/AlGaAs quantum well 
by using resistive detection. 
This detection method provides the high sensitivity and selectivity of nuclear spins in the well, 
and thus, we clearly observed the optical nuclear polarization spectra for $\sigma^+$ and $\sigma^-$ excitations
and the dependences of the optical nuclear polarization on the laser power and pumping time. 
We constructed a theoretical formulation, including the steady state electron spin polarization for the modulation doped system under optical pumping,  
to describe the nuclear magnetic field induced by optical pumping in the quantum Hall system.  
The optical nuclear polarization spectra directly reflect the absorption spectra, 
and the sign of the nuclear magnetic field coincides with the electron spin state within the lowest Landau level interband transitions. 
This makes it possible to apply spectroscopy mediated by optical nuclear polarization  
as a novel method for electron-spin-resolved spectroscopy.   
We found that 
the phonon emission process, which usually assists the nuclear polarization due to the energy conservation,  
influences the magnitude of the optical nuclear polarization through an increase in temperature. 
We also discussed that
the optical nuclear polarization by pumping the lowest energy level can be accomplished by the electron-electron interaction. 
The observation of the change in the electron-nuclear cross relaxation rate under optical pumping 
provides the possibility 
to further control the nuclear spin polarization by tuning the pumping time and intensity profile for laser illumination. 

The information about the optical nuclear polarization we presented here 
will be valuable when applying
the conventional, optically detected, and resistively detected NMR 
to a small ensemble of the nuclear spins in microscopic samples and various materials. 
Furthermore, 
since nuclear spins in the semiconductor system 
are good candidates for qubits in solid-state quantum information technology, 
the effective manipulation of the nuclear spins that is 
based on the detailed mechanism of the optical nuclear polarization 
leads to rich quantum information processing.

\acknowledgments
The authors are grateful to G. Yusa, J. Hayakawa, S. Watanabe, and T. Hatano 
for their fruitful discussions,
and K. Muraki for providing high quality wafers. 
One of the authors (TY) was supported by the JSPS Research Fellowship for Young Scientist (No. 24-1112).


\begin{thebibliography}{99}
\bibitem{noteDNP}
The current induced nuclear polarization was not observed in our sample under these conditions.
\bibitem{check}
We checked that the parallel conduction did not affect $V_{peak}$.
\bibitem{jouken}
To obtain the upper (lower) curve, the optical pumping was performed 
with $\lambda=809$~nm, $P=1.6$~W/cm$^2$, and $\sigma^{+}$ ($\sigma^{-}$) 
at $\nu= 0.62$ for $\tau_{pump}=150$~s.
\bibitem{Temp}
In this discussion, we incorporate the effect of the temperature increase 
into the changes in $\langle S _z\rangle_{eq}$ and $\tau_e/T_{1e}$ in Eq.~(\ref{DSz}). 
There is a possibility, however, 
that Eq.~(\ref{DSz}) is not appropriate for describing 
the electron spin polarization in such a situation.
\bibitem{Coulomb} 
In the $\sigma^{+}$ excitation (pumping the upper Zeeman level), 
the electron-electron interaction can also play an important role to polarize nuclear spins.  
Even though this interaction is mainly attributed to the nuclear polarization, 
the phonon emission occurs in the electron relaxation from the upper Zeeman level.
\end{thebibliography}

\end{document}